\documentclass{aa}
\usepackage{txfonts,natbib,graphicx}
\bibpunct{(}{)}{;}{a}{}{,}

\begin{document}

\title{Australia Telescope Compact Array imaging of circumstellar
HCN line emission from R Scl} 

\author{T. Wong\inst{1} \and F.~L. Sch\"{o}ier\inst{2,3} \and
M. Lindqvist\inst{4} \and H. Olofsson\inst{3}}

\institute{
CSIRO Australia Telescope National Facility, PO Box 76, 
Epping NSW 1710, Australia
\and
Sterrewacht Leiden, PO Box 9513, 2300 RA Leiden, The Netherlands
\and
Stockholm Observatory, AlbaNova, 106 91 Stockholm, Sweden
\and 
Onsala Space Observatory, 439 92 Onsala, Sweden}

\offprints{T. Wong, \email{Tony.Wong@csiro.au}}

\date{Received / Accepted}

\abstract{
We present radio-interferometric observations of HCN $J=1\rightarrow0$
line emission from the carbon star R Scl, obtained with the interim
3-mm receivers of the Australia Telescope Compact Array.  The emission
is resolved into a central source with a Gaussian $FWHM$ of
$\sim$1\arcsec, which we identify as the present mass loss envelope.
Using a simple photodissociation model and constraints from
single-dish HCN spectra, we argue that the present mass-loss rate is
low, $\sim 2 \times 10^{-7}$ M$_\odot$~yr$^{-1}$, supporting the idea
that R Scl had to experience a brief episode of intense mass loss in
order to produce the detached CO shell at $\sim$10\arcsec\ radius
inferred from single-dish observations.  Detailed radiative transfer
modelling yields an abundance of HCN relative to H$_2$,
$f_{\mathrm{HCN}}$, of $\sim10^{-5}$ in the present-day wind.  There
appears to be a discrepancy between model results obtained with
higher transition single-dish data included and those from the
$J=1\rightarrow0$ interferometer data alone, in that the
interferometer data suggest a smaller envelope size and larger HCN
abundance than the single-dish data.  The lack of HCN in the detached
shell, $f_{\mathrm{HCN}}\lesssim 2\times 10^{-7}$, is consistent with
the rapid photodissociation of HCN into CN as it expands away from the
star.
}

\titlerunning{ATCA imaging of HCN emission from R Scl}

\maketitle

\section{Introduction}

In the late stages of asymptotic giant branch (AGB) evolution,
material is ejected via a slow (5--30 km~s$^{-1}$) stellar wind into
a circumstellar envelope of gas and dust.  This mass loss (on the
order of $10^{-7}$--$10^{-5}$ M$_{\odot}$~yr$^{-1}$) is a key process
for enriching and replenishing the interstellar medium, yet the
mechanisms by which it occurs are not understood in detail.  In
particular, there is evidence that strong variations in the mass loss
rate, perhaps related to flashes of helium shell burning, can lead to
the formation of circumstellar detached shells \citep{Olofsson:90}.
Such shells can be inferred from their excess emission at long
infrared wavelengths (due to a lack of hot dust close to the star,
\citealt{Willems:88}) and confirmed by a ring-like morphology in maps
of dust or CO line emission.

Detached shells have been observed around about a half-dozen AGB stars
\citep[see review by][]{Wallerstein:98}, but only \object{U Cam}
\citep{Lindqvist:96,Lindqvist:99} and \object{TT Cyg}
\citep{Olofsson:98,Olofsson:00} have been mapped with millimetre-wave
interferometers.  Both cases reveal a beautifully symmetric shell
showing uniform expansion from the star.  High-resolution molecular
line observations offer the possibility not only to confirm
detachment, but also to study variations in mass loss on short ($<$100
yr) timescales, related to the shell thickness, and to look for
departures from spherical symmetry.

As one of the initial targets for the 3-mm receiver systems being
installed on the Australia Telescope Compact Array (ATCA), we have
imaged the HCN $J=1 \rightarrow 0$ emission from the carbon star \object{R
Sculptoris} (R Scl).  Previous observations by \citet{Olofsson:96} with
the Swedish-ESO Submillimetre Telescope (SEST) have indicated a
detached shell which is only marginally resolved (radius
$\approx$10\arcsec) in the CO $J=3\rightarrow2$ line, suggesting a
relatively young age for the shell ($\lesssim 10^3$\,yr, using the
observed shell expansion velocity of 16.5 km~s$^{-1}$ and a distance of 360
pc; \citealt{SchoierOlofsson2001}).  Detections of other molecules like
HCN, CS, and CN have also been reported by \citet{Olofsson:96}.  These
molecules are more easily photodissociated than CO, but it is not
clear whether they are located in the detached shell or confined to
the current mass loss envelope, which the SEST data, with limited
spatial resolution, could not distinguish from the shell.  Another
outstanding question is the relationship between the molecular gas
shell and the larger dust shell (radius $\approx$20\arcsec) seen in
scattered optical light \citep{Gonzalez:01,Delgado03a}.  The recent
commissioning of a southern millimetre array motivated us to obtain
much higher resolution data toward this object.

In this paper we present the results of our ATCA observations and 
analyse them, along with new and previously published single-dish
spectra, using a detailed radiative transfer code for modelling 
circumstellar line emission \citep{Lindqvist00,SchoierOlofsson2001}.

\section{Observations and data reduction}\label{sec:obs}


\begin{figure}
\resizebox{\hsize}{!}{
\includegraphics[clip=true,viewport=0 0 384 224]{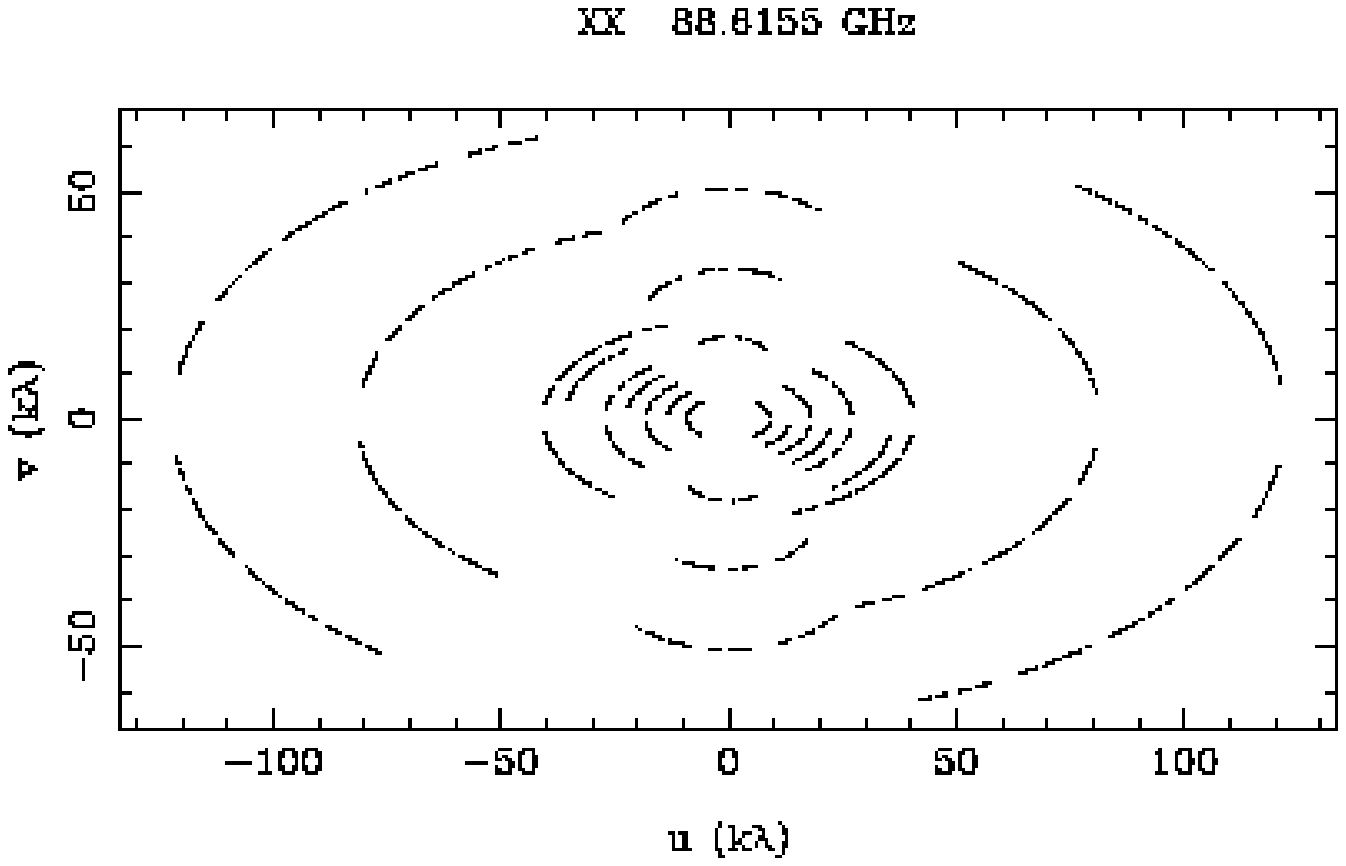}}
\caption{
Coverage of the visibility plane obtained after combining observations
in all four ATCA configurations (EW214, EW352, H168, and 750A).  
\label{fig:uvcov}}
\end{figure}

\subsection{Interferometer data}\label{sec:obsat}

At the time of the observations, the ATCA had three antennas of 22~m
diameter equipped with dual polarisation 3-mm receivers covering the
bands 84.9--87.3 and 88.5--91.3 GHz.  We observed R Scl between 2002
June and October in four different array configurations: EW214
(baselines of 31, 61, and 92 m), EW352 (46, 77, and 122 m), H168 (61,
111, and 171 m), and 750A (138, 275, and 413 m).  Note that in the
H168 configuration, the 3-mm antennas are situated in a north-south
line; the remaining configurations are east-west arrays.  The 750A
observations were taken over two nights spaced 3 days apart;
observations in the other configurations occurred on single nights.
The combined visibility plane coverage is shown in
Fig.~\ref{fig:uvcov}.

All observations were conducted in mostly clear weather, with
above-atmosphere single-sideband system temperatures of $T_{\rm sys}^*
\approx 300$~K near the zenith.  The correlator was configured to
receive both linear polarisations in two frequency windows: a
narrowband (spectral line) window centred on the HCN $J=1\rightarrow0$
line at 88.6 GHz with 128 channels across 32 MHz, and a wideband
(continuum) window centred at 89 GHz with 32 channels across 128
MHz. The pointing and phase centre was at $\alpha_{2000} = 01^{\rm h}
26^{\rm m} 58\fs49$, $\delta_{2000} = -32\degr 32\arcmin 35\farcs5$,
which is 5\arcsec\ east of the stellar position (in the subsequent
data analysis we have applied an offset so that the phase centre
coincides with the star).  At 89 GHz the ATCA primary beam has a $FWHM$
of about 35\arcsec.

Gain calibration was performed by frequent observations (every 10--20
minutes) of the nearby quasars B0104$-$408 or B0208$-$512.  The amplitude
and phase gains were derived using the continuum window data and then
transferred to the spectral line window after applying offsets
determined from a half-hour integration on the bandpass calibrator
(B1921$-$293, B2255$-$282, or B0420$-$014).  Channel-dependent gains were
also determined from the bandpass calibrator.  Uranus was used to set
the flux scale, assuming a uniform disk with brightness temperature of
134~K.  We also adjusted the antenna pointing once an hour on
B0208$-$512; typical pointing shifts were $\sim$5\arcsec.

All data processing was conducted using the MIRIAD package.  As an
initial step in the reduction, values of $T_{\rm sys}^*$ measured via
the chopper wheel method \citep{Kutner:81} were used to rescale the
data, then an elevation-dependent gain curve was applied based on
observations of an SiO maser taken on 2002 June 4.  Subsequent
processing was as described above using standard routines.  The
calibrated visibilities were Fourier transformed using robust
weighting and a 0\farcs5 pixel size with a channel spacing of 1
km~s$^{-1}$ (roughly the effective velocity resolution given the
original channel spacing of 0.25 MHz $\approx 0.8$ km~s$^{-1}$).  The
maps were then cleaned down to a 2$\sigma$ level over the inner
20\arcsec\ $\times$ 20\arcsec.

Over the course of the observations, it became clear that there were
residual phase errors in the data that were not being removed by the
standard technique of phase referencing off a nearby quasar.  These
phase errors, revealed as phase jumps when switching sources, can be
modelled as errors in the antenna positions of up to a few mm.  For
most of the data, we were able to minimise the effects of these phase
errors by refining the baseline solution from observations of multiple
quasars taken on the same or adjacent days.  No such correction could
be applied to the EW352 data, which are therefore excluded from the
image analysis below (Sect.~\ref{sec:image}), but are included in the
visibility analysis after self-calibration (Sect.~\ref{sec:vis}).

\begin{table}
\caption[]{Single-dish HCN molecular line observations of
\object{R~Scl}.}\label{radio}
\begin{tabular}{llcccccc}
\hline
\hline
\multicolumn{1}{c}{Trans.} & 
\multicolumn{1}{c}{Tel.} &
\multicolumn{1}{c}{${\int T_{\mathrm{mb}} dV} ^{\mathrm a}$} &
\multicolumn{1}{c}{${T_{\mathrm{mb}}} ^{\mathrm b}$} &
\multicolumn{1}{c}{$FWHM ^{\mathrm b}$} &
\multicolumn{1}{c}{${V_{\star}} ^{\mathrm b}$} \\
& &
\multicolumn{1}{c}{$[\mathrm{K\,km\,s}^{-1}]$} &
\multicolumn{1}{c}{$[\mathrm{K}]$} &
\multicolumn{1}{c}{$[\mathrm{km\,s}^{-1}]$} &
\multicolumn{1}{c}{$[\mathrm{km\,s}^{-1}]$} \\
\hline
$J=1\rightarrow0$ & SEST     & $1.6$ & $0.08$ & $19.7$ & $-16.2$ \\
$J=3\rightarrow2$ & SEST     & $5.4$ & $0.36$ & $14.6$ & $-17.4$ \\
$J=3\rightarrow2$ & HHT$^{\mathrm c}$	& $2.4$ & $0.13$ & $16.1$ & $-18.5$ \\
$J=4\rightarrow3$ & HHT$^{\mathrm c}$	& $4.1$ & $0.29$ & $14.8$ & $-17.7$ \\
\hline
\end{tabular}
\noindent$^{\mathrm a}$ Total integrated intensity calculated over full extent
of line. \\
\noindent$^{\mathrm b}$ Estimated from a Gaussian fit to the observed
spectrum. \\
$^{\mathrm c}$ From \citet{Bieging01}. \\
\end{table}

\subsection{Single-dish data}\label{sec:obssd}

As a complement to the interferometer observations, multi-transition
single-dish data have been obtained.  The single-dish data are used to
check that the interferometer recovers all of the source flux and also
to constrain the circumstellar HCN models.  Using the SEST, the HCN
$J=1\rightarrow0$ and $J=3\rightarrow2$ transitions were observed in
February 2003 and December 1996 respectively.  Two acousto-optical
spectrometers were used at SEST (86~MHz bandwidth with 43~kHz channel
separation, and 1~GHz bandwidth with 0.7~MHz channel separation).
Dual beam switching (beam throws of about 11\arcmin), in which the
source was placed alternately in the two beams, was used to eliminate
baseline ripples.  The line intensities are given in the main beam
brightness temperature scale ($T_{\rm mb}$), i.e., the antenna
temperature has been corrected for atmospheric attenuation (using the
chopper wheel method) and divided by the main beam efficiency.  In
addition, recently published HCN $J=3\rightarrow2$ and
$J=4\rightarrow3$ spectra from \citet{Bieging01}, obtained with the
Heinrich Hertz Submillimeter Telescope (HHT), have been used. 
The single-dish data are presented in Table~\ref{radio} and 
Sect.~{\ref{sec:hcnmod}.


\begin{figure*}
\centering
\includegraphics[width=15cm]{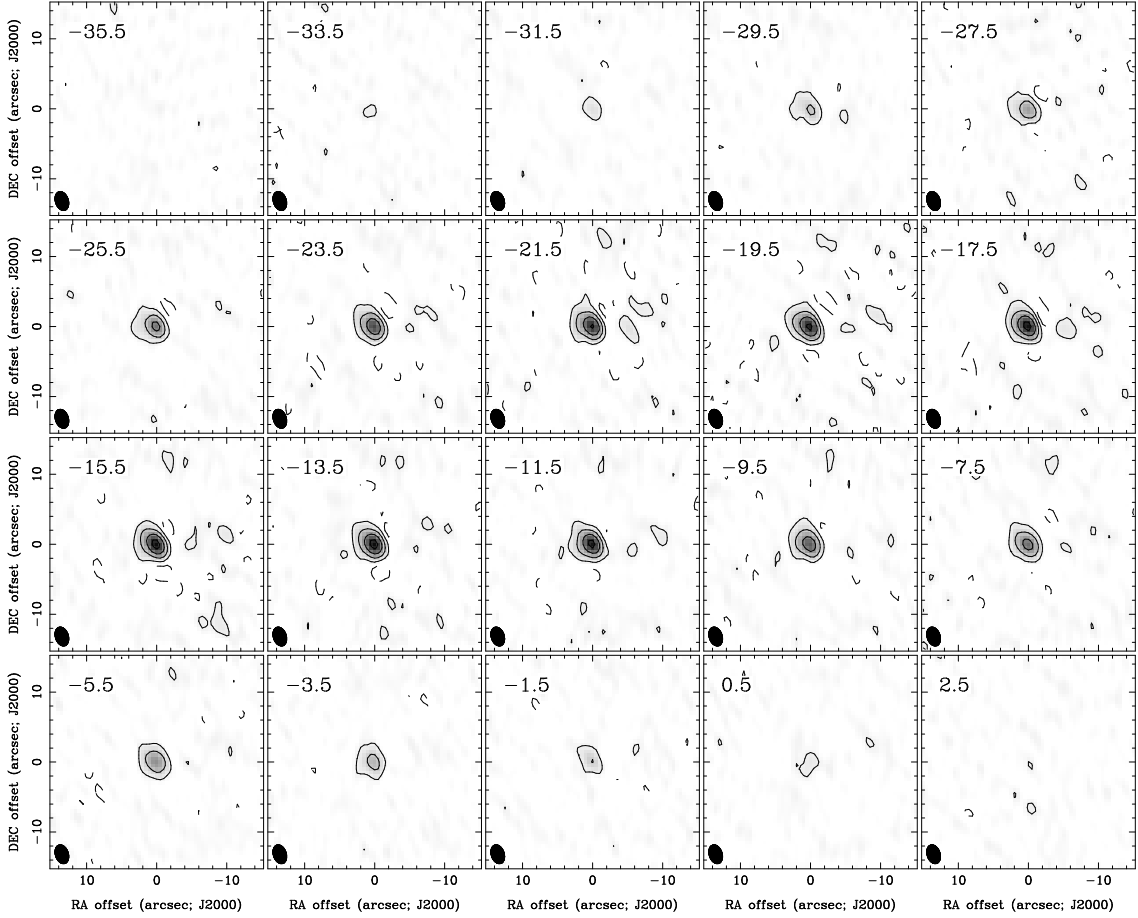}
\caption{
Deconvolved channel maps of HCN emission from R Scl.  The contour
levels are 0.05$n^2$ Jy~beam$^{-1}$, where $n=1,2,3,4$, and the beam size
is 2\farcs7 $\times$ 1\farcs8.  The velocity channels (given in the
LSR frame) have been binned to 2 km~s$^{-1}$.
\label{fig:chanmaps}}
\end{figure*}

\section{Observational results}

\subsection{Channel maps}\label{sec:image}

Figure~\ref{fig:chanmaps} shows the HCN $J=1\rightarrow0$ channel maps
(with the EW352 data excluded) binned into 2 km~s$^{-1}$ channels.  The
synthesised beam is 2\farcs7 $\times$ 1\farcs8 with a position angle
of 18\degr.  The RMS noise in the 1 km~s$^{-1}$ channel maps is $\sigma$ =
23 mJy~beam$^{-1}$; the peak emission reaches a level of about 40$\sigma$.  We
note here that the HCN $J=1\rightarrow0$ line actually consists of
three hyperfine components of relative strength 3:5:1 (in order of
increasing frequency), with the low-frequency satellite line shifted
$+$4.8 km~s$^{-1}$ and the high-frequency satellite line shifted $-$7.1
km~s$^{-1}$ with respect to the main line.


\begin{figure}
\begin{center}
\resizebox{\hsize}{!}{\includegraphics{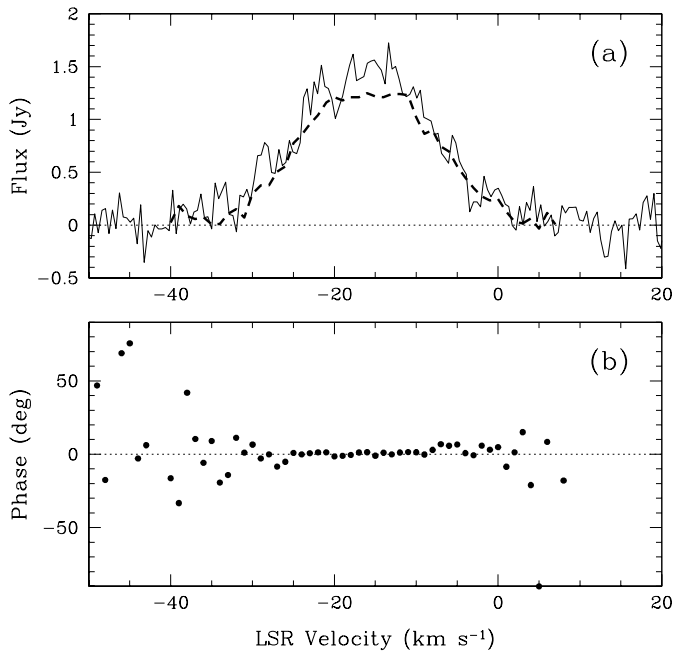}}
\caption{
(a) HCN (1--0) spectra for R Scl, with the dashed line derived from
the ATCA image (integrating the channel maps over a 6\arcsec\ $\times$
6\arcsec\ box centred on $\alpha_{2000}$=$01^{\rm h} 26^{\rm m}
58\fs09$, $\delta_{2000}$=$-32\degr 32\arcmin 35\farcs5$) and the thin
solid line from the SEST observations. (b) Average visibility phase
spectrum, including data from all baselines and array configurations.
The phase vs.\ time has been self-calibrated to bring the average
phase to $0\degr$.
\label{fig:totspec}}
\end{center}
\end{figure}

Inspection of the channel maps suggests that the source is moderately
resolved, a result which is confirmed by the visibility analysis in
Sect.~\ref{sec:vis}.  The source also appears to be somewhat extended to
the east.  We believe that this extension is most likely attributable
to the instrumental phase errors discussed in Sect.~\ref{sec:obsat}: mapping
the EW214 data alone shows the emission to be displaced by about
0\farcs6-0\farcs7 to the east of the star's position, whereas the
other tracks do not show such an offset.  The phase error needed to
produce such an offset is $\sim$30\degr, which is within the range
expected from residual errors in the baseline solution.

Fitting two-dimensional Gaussians to the individual channel maps, we
find no systematic variations in source size or position angle with
velocity.  Across the peak of the line profile, from $-10$ to $-20$
km~s$^{-1}$, the average values of the major and minor axis $FWHM$ are
3\farcs1 and 2\farcs3 respectively, with a position angle of 36\degr\
(note that the beam has not been deconvolved from these values).

We see no indication of an expanding shell structure corresponding to
that seen in the SEST CO $J=3\rightarrow2$ data.  A shell of radius
10\arcsec\ would appear well within the half-power radius of the
primary beam, especially on the eastern side, since the pointing
centre was offset 5\arcsec\ east of the star.  Moreover, an {\it
expanding} spherical shell would become smaller near the extreme
velocities (aside from some complications due to the hyperfine
structure of the HCN line), and thus certainly fall within the primary
beam, yet there is no indication that the source structure or size
changes with velocity.

\subsection{Spectra}\label{sec:spectra}

The integrated HCN $J=1 \rightarrow 0$ ATCA spectrum is shown in
Fig.~\ref{fig:totspec}(a), overlaid on the equivalent SEST spectrum.
A Gaussian fit to the ATCA spectrum gives a mean velocity, $V_{\star}$,
 of $-15.8$
km~s$^{-1}$ and a $FWHM$ of 19.1 km~s$^{-1}$; however, this may be
affected by the hyperfine structure of the HCN line.  The peak flux of
1.2~Jy agrees, within the calibration uncertainties, with the
single-dish flux of 1.4~Jy detected with the SEST ($S = \eta_B T_{\rm
mb} \Gamma^{-1}$, where $\eta_B$=0.75, $T_{\rm mb}$=75~mK, and
$\Gamma^{-1}$=25~Jy~K$^{-1}$).  Thus, our lack of interferometer
spacings less than 30~m does not appear to have resulted in a
significant amount of ``missed'' flux, at least within the 58\arcsec\
SEST beam.  This confirms that virtually all of the HCN
emission comes from the central component.

The higher-transition spectra taken with the SEST and HHT (see
Table~\ref{radio}) are in general agreement with the $J = 1 \rightarrow 0$
spectra, except that their mean velocity is $\approx-17.5$
km~s$^{-1}$ and the linewidths are somewhat narrower,
$FWHM\approx15$~km~s$^{-1}$.  Since these lines are less affected by
hyperfine structure, they should serve as a better indication of the
kinematics of the envelope.  The spectra are presented and discussed
in Sect.~\ref{sec:hcnmod}.


\begin{figure}
\resizebox{\hsize}{!}{\includegraphics{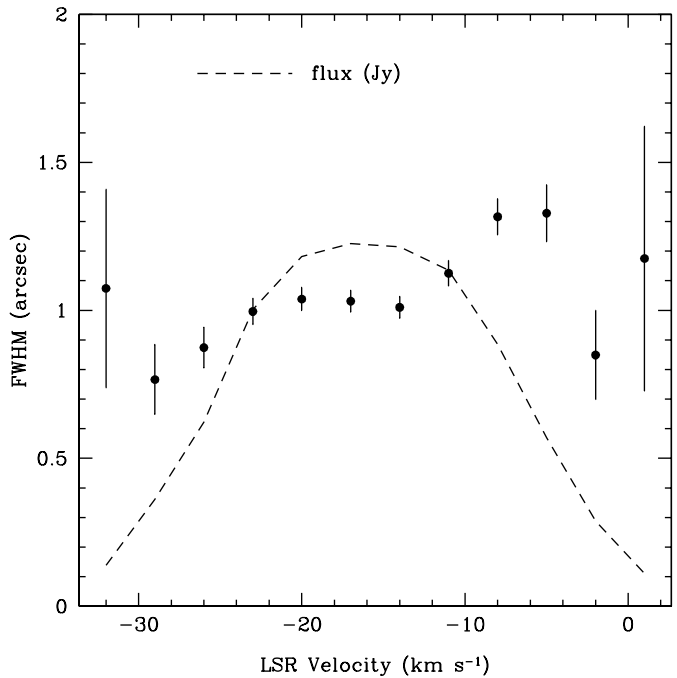}}
\caption{
{\it FWHM} of axisymmetric Gaussian fits to the visibility data averaged in 3
km~s$^{-1}$ channels.  The source is constrained to lie at the phase centre.
The flux of the model Gaussian, on the same numerical scale but in Jy, 
is overplotted as a dashed line.
\label{fig:uvfit}}
\end{figure}


\begin{figure}
\resizebox{\hsize}{!}{\includegraphics{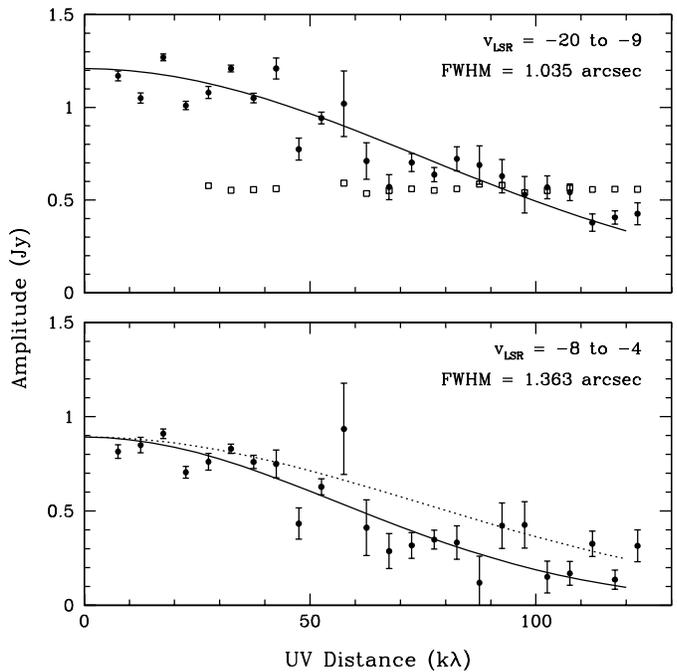}}
\caption{
{\it (Top)} Averaged visibility amplitudes, in 5 k$\lambda$ bins,
across the peak of the line profile (filled circles).  The solid line
is the model fit, with a $FWHM$ of 1\farcs0.  The open squares represent
the averaged visibility amplitudes of the gain calibrator, B0104$-$408.
{\it (Bottom)} Same but for the high-velocity edge of the line profile.  The
solid line is the model fit, with a $FWHM$ of 1\farcs4, whereas the dotted
line corresponds to a $FWHM$ of 1\farcs0.
\label{fig:uvmod}}
\end{figure}

\subsection{Visibility analysis}\label{sec:vis}

The results of Sect.~\ref{sec:image} indicate that the HCN
$J=1\rightarrow0$ emission comes from a compact central source that is
only moderately resolved by our observations.  We now proceed to
analyse the visibilities directly, under the assumption that the
emission structure is axisymmetric, i.e. there is no phase 
information in the visibilities.  
This assumption allows us to self-calibrate the data
in phase (in practice a 30-minute interval was used) and therefore
include the EW352 data, which were strongly affected by phase errors.
We stress that applying self-calibration with only three
interferometer baselines is guaranteed to produce an axially
symmetric source at the phase centre, as long as closure errors are
negligible.  Independent evidence that there is little additional
source structure comes from the averaged visibility phase spectrum
shown in Fig.~\ref{fig:totspec}(b).  Since the channel-to-channel
phase has {\it not} been self-calibrated, but only calibrated using
the bandpass calibrator, the flatness of the phase spectrum
indicates that the source structure is the same at all velocities.

We fit a model of an axisymmetric Gaussian to the
self-calibrated visibilities.  The resulting $FWHM$ in arcseconds is
plotted in Fig.~\ref{fig:uvfit}, along with the flux of the model.
Most of the values are close to 1\farcs0, although somewhat higher
values (1\farcs3) are seen at the redshifted edge of the line.  An
average over all points gives 1\farcs05 $\pm$ 0\farcs17.  Note that a
$FWHM$ of 1\arcsec\ corresponds to a radius at half maximum, at the
distance of R~Scl, of about 200 AU ($3 \times 10^{15}$ cm).  As with
the image analysis in Sect.~\ref{sec:image}, we do not see the
variation in size with line-of-sight velocity expected for an
expanding shell.

The top panel of Fig.~\ref{fig:uvmod} shows a 1\farcs0 Gaussian
model fit to the peak of the line profile, overlaid on the visibility
amplitude averaged in 5 k$\lambda$ bins.  For comparison, the open
squares give the average visibility amplitude for the phase
calibrator, B0104$-$408, during the two October runs in 750A (the
amplitude was $\sim$30\% lower in the June runs, but this may be due
to intrinsic variability of this quasar and/or errors in relative flux
calibration).  Note that the amplitude of the calibrator does not show
a significant decrease with baseline length, as would be expected if
atmospheric phase decorrelation were affecting the averaged visibility
amplitudes.  Consequently, there is no doubt that the HCN emission
from R Scl has been spatially resolved by our observations.
However, because our baselines are not long enough to show a 
possible null in the visibility function, other types of extended
models besides a Gaussian (e.g., a uniform disk or optically thin
sphere) are not excluded.

The bottom panel of Fig.~\ref{fig:uvmod} shows a 1\farcs36 Gaussian
model fit to the redshifted edge of the line profile, which showed
indications of a larger source size in Fig.~\ref{fig:uvfit}.  While
the fit appears to be a reasonable one, there are clearly some bins
that agree better with the 1\arcsec\ source size derived from the full
velocity range (dotted line).  Thus, the significance of the larger
source size at these velocities appears doubtful.  We note that for an
expanding shell that has been well-resolved, we would expect the
source size to decrease, rather than increase, near the line wings.

\section{Radiative transfer modelling}

To model the circumstellar line emission we have used a detailed
non-LTE radiative transfer code, based on the Monte Carlo method
\citep{Bernes79}.  The code has been described in detail in
\citet{Schoier2000} and \citet{SchoierOlofsson2001} and has been
benchmarked against a wide variety of molecular line radiative
transfer codes in \citet{Zadelhoff_etal2002}.  The code assumes a
spherical envelope expanding at constant velocity and includes
radiative excitation through vibrationally excited states, and a full
treatment of line overlaps between various hyperfine components.  Our
procedure and the relevant molecular data are summarised in
\citet{Lindqvist00}.

From the HCN line emission alone it is not possible to constrain both
the abundance of HCN (relative to H$_2$) and the mass loss rate.  For
our analysis, therefore, we first adopt a mass loss rate for the
present day wind.  We construct a model of the CO envelope to derive
the temperature profile, then determine the best-fit values for the
HCN abundance and envelope size, using both the single-dish and
interferometer data as constraints.  The HCN abundance,
$f_{\mathrm{HCN}}$, is taken to be fixed throughout the envelope out
to a specified outer radius, $r_{\mathrm{HCN}}$.

\subsection{The mass loss rate and temperature profile}

\citet{LeBertre97} estimated the present mass loss rate for R~Scl to
be approximately $10^{-7}$ M$_{\odot}$~yr$^{-1}$, scaled to our
adopted distance of 360~pc, based on radiative transfer modelling of
the observed dust emission.  This method assumes that the dust grain
properties and the dust-to-gas mass and velocity ratios in R~Scl are
the same as in \object{IRC+10216}, for which the mass loss rate is
taken to be $2.5 \times 10^{-5}$ M$_{\odot}$~yr$^{-1}$ at a distance
of 200 pc.  A larger mass loss rate of $4 \times 10^{-7}$
M$_{\odot}$~yr$^{-1}$ (again scaled to our adopted distance) is
derived by \citet{Gustafsson97}, using \ion{K}{i} scattered light
observations, but this estimate must be considered relatively
uncertain since it is based on a method which has not been extensively
tested.  As long as the size of the emitting region is limited by
photodissociation rather than excitation, a rough estimate of the mass
loss rate can also be obtained from the observed size of the HCN
brightness distribution, using a simple photodissociation model
\citep[cf.,][]{Lindqvist00}.  For an estimated expansion velocity of
10.5 km~s$^{-1}$ (see below), a mass loss rate of approximately $5
\times 10^{-7}$ M$_{\odot}$~yr$^{-1}$ is derived assuming the same
dust properties as for IRC+10216 \citep{SchoierOlofsson2001}.  Based
on these estimates, we adopt a mass loss rate of $2\times 10^{-7}$
M$_{\odot}$~yr$^{-1}$ for the present epoch.  Thus, the mass loss rate
for R~Scl appears higher than the present mass loss rates of sources
with relatively old detached shells like TT~Cyg \citep[$\dot{M}\sim
3\times 10^{-8}$~M$_{\odot}$\,yr$^{-1}$;][]{Olofsson:98,Olofsson:00}
but comparable to that of the `young shell' source U~Cam
\citep[$\dot{M}\sim 2.5\times
10^{-7}$~M$_{\odot}$\,yr$^{-1}$;][]{Lindqvist:96,Lindqvist:99}.

The expansion velocity of the wind is a free parameter in the
radiative transfer that is well constrained by the width of the
observed lines.  A value of 10.5 km~s$^{-1}$ can reproduce both the
$J=1\rightarrow 0$ line and the higher $J$ transition HCN line
profiles, which are much less affected by hyperfine structure (see
Sect.~\ref{sec:hcnmod}).  The source is assumed to have a distance of
360~pc and the stellar radiation to be described by a blackbody with a
luminosity of 5500~L$_{\odot}$ and a temperature of 2500~K \citep[][
Kerschbaum priv. comm.]{SchoierOlofsson2001}.
The inner radius of the envelope is fixed at $3 \times 10^{14}$~cm (20 AU). 

Using the above parameters we have computed a CO model of the
envelope, adopting a CO abundance of $1\times 10^{-3}$ relative to
H$_2$.  The outer radius of the CO envelope, $\sim$$3 \times
10^{16}$~cm (6\arcsec), is obtained from a CO photodissociation model
\citep{Mamon88}.  In addition to determining the steady-state level
populations of the CO molecule, the code simultaneously solves the
energy balance equation, including the most relevant heating and
cooling processes.  Heating is dominated by collisions between the
dust and gas, while cooling is generally dominated by CO line
emission, although adiabatic cooling due to the expansion of the
envelope is also important.  We have assumed that R~Scl has the same
dust properties as IRC+10216.  Our excitation analysis allows for a
self-consistent treatment of molecular line cooling.  In this way the
kinetic temperature structure of the envelope is determined, for use
in the subsequent HCN line modelling.  Throughout most of the envelope
the temperature structure is well described, to within 10\%, by a
simple power-law $T_{\mathrm{kin}}=T_0(r_0/r)^{0.47}$, where
$T_0=182$~K and $r_0=1\times10^{16}$~cm.  However, in the inner parts
of the envelope ($r\leq 1\times10^{15}$~cm) the temperature rises more
dramatically to a value of 1350~K at the inner edge of the envelope. 

We find that our derived CO line intensities are much lower than the
observed ones, consistent with most of the observed CO line emission
coming from a detached shell distinct from the present mass loss
envelope.  A more thorough discussion will be presented in Lindqvist
et al. (in prep.).


\begin{figure}
\resizebox{\hsize}{!}{\includegraphics[angle=-90]{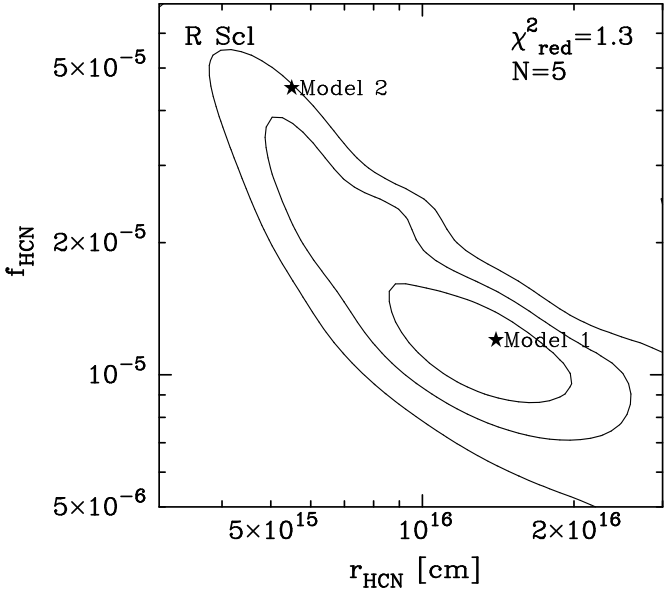}}
\caption{$\chi^2$ map showing the sensitivity of the HCN envelope
model to the adjustable parameters $f_{\mathrm{HCN}}$ and
$r_{\mathrm{HCN}}$.  Contours are drawn at $\chi^2_{\mathrm{min}} +
(2.3, 6.2, 11.8)$ indicating the 68\% (`1$\sigma$'), 95\%
(`2$\sigma$'), and 99.7\% (`3$\sigma$') confidence levels,
respectively.  The quality of the best fit model (Model 1) can be
estimated from the reduced chi-squared statistic
$\chi^{2}_{\mathrm{red}}$=$\chi^{2}_{\mathrm{min}}$/($N$$-$2), and is
shown in the upper right corner. Also shown is the number of
observational constraints used, $N$, which includes the single-dish
spectra and the ATCA spectrum at the centre pixel.  The model that
best reproduces the visibility amplitudes obtained by ATCA (Model 2)
is also shown for comparison.
\label{fig:chi2}}
\end{figure}


\begin{figure*}
\centering
\includegraphics[width=17cm]{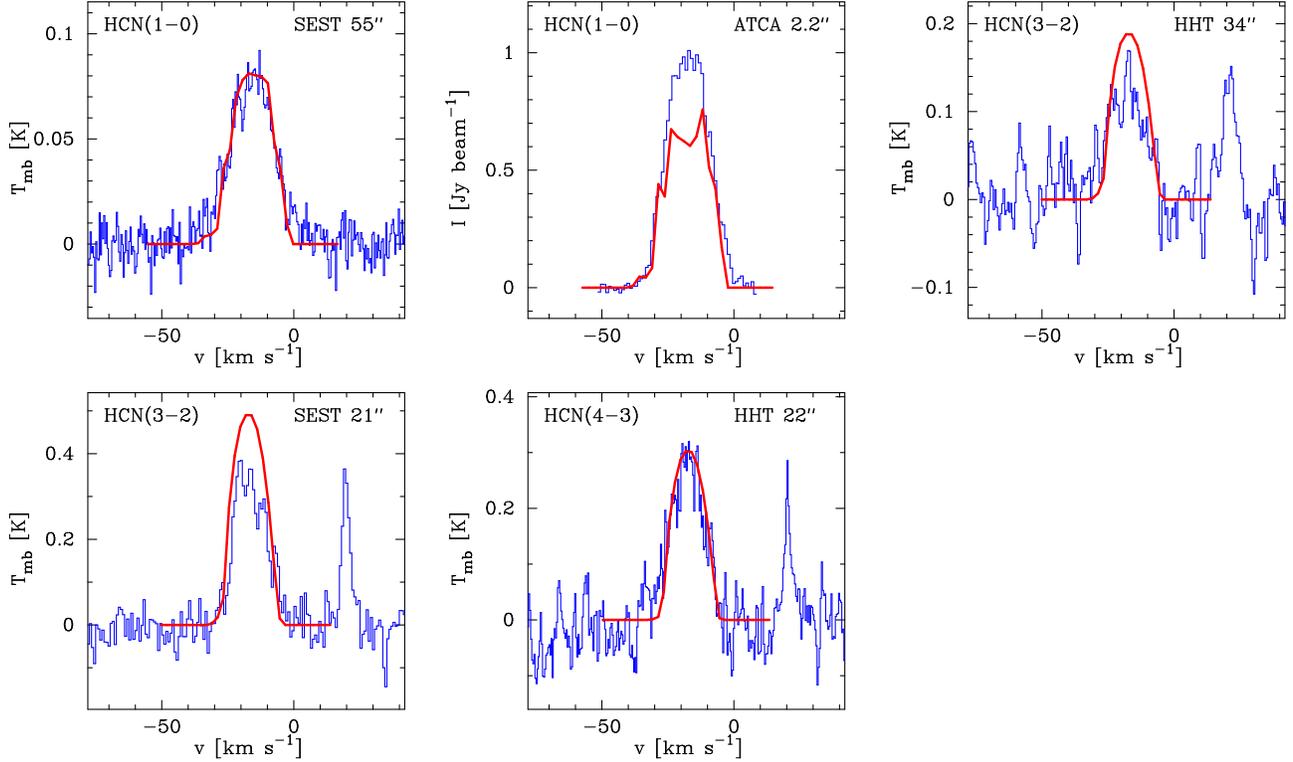}
\caption{Best fit model (full line), using a HCN abundance of $1.2
\times 10^{-5}$ and envelope size of $1.4 \times 10^{16}$~cm, overlaid
on the observed spectra (histograms).  The narrow spectral features in
the $J=3\rightarrow2$ and $J=4\rightarrow3$ spectra are from vibrationally excited HCN.  The
ATCA spectrum is taken at the centre pixel of the image cube.
\label{fig:overlay}}
\end{figure*}

\subsection{HCN line modelling}\label{sec:hcnmod}

With the physical structure of the envelope fixed, a set of models was
calculated where the HCN abundance relative to H$_2$,
$f_{\mathrm{HCN}}$, was allowed to vary from $5 \times 10^{-6}$ to $5
\times 10^{-5}$ and the outer radius of the HCN-emitting envelope from
$2 \times 10^{15}$~cm to $5 \times 10^{16}$~cm.  For each model a
total $\chi^2$ \citep[for a definition see][]{SchoierOlofsson2001} was
determined via comparison with the observed spectra, assuming a
calibration uncertainty of 20\% for each spectrum.  In addition to the
single-dish data presented in Table~\ref{radio}, the ATCA spectrum at
the centre pixel of the maps was used in the $\chi^2$ analysis.  Note
that the $\chi^2$ map (Fig.~\ref{fig:chi2}) shows that the line
intensities become relatively insensitive to the adopted size of the
HCN envelope when $\gtrsim 3\times 10^{16}$~cm.  This is where the
emitting region becomes excitation limited.  Such solutions, however,
are implausible given that the CO envelope is calculated to extend
only to $3 \times 10^{16}$~cm, and CO is less readily photodissociated
than HCN.
The best fit model has an abundance of $(1.2 \pm 0.4) \times 10^{-5}$
and a size of $(1.4 \pm 0.5) \times 10^{16}$~cm (Fig.~\ref{fig:chi2};
Model 1).  The reduced $\chi^2$ has a value of 1.3, indicating a good
fit.  However, as can be seen by comparing this model with the
observed spectra (Fig.~\ref{fig:overlay}), the model ATCA spectrum is
too weak and presents more of a double-peaked rather than parabolic
line shape.  
Note that we have not included the
vibrationally excited lines seen in the $J=3\rightarrow2$ and
$J=4\rightarrow3$ spectra in Fig.~\ref{fig:overlay} in our analysis,
given that they are masing and highly variable \citep{Bieging01}.

The main source of the discrepancy between our model and
the ATCA data is that the size of the model envelope is approximately
five times larger than that obtained from the interferometer
$uv$-plane data (Sect.~\ref{sec:vis}).  Applying the ATCA $uv$
sampling to the model envelope, we find that the model severely
underpredicts the flux observed on long baselines
(Fig.~\ref{fig:uvmodel}).  As shown by the dashed line, a better fit
is obtained by raising the abundance of HCN ($f_{\mathrm{HCN}} =
4.5\times 10^{-5}$) while making the envelope smaller
($r_{\mathrm{HCN}} = 5.5\times 10^{15}$~cm) (Model 2).  However, such
a model would predict much stronger emission (by $\sim$50\%) in the
higher $J$ transitions, in conflict with the single-dish constraints
(Fig.~\ref{fig:chi2}).  We also note that Model 2 predicts a decrease
in source size at the extreme velocities of the line profile, which,
while expected for a spherically symmetric wind expanding at a
constant velocity, is not evident in the observations
(Fig.~\ref{fig:uvfit}).

It is not clear how to resolve the discrepancy between the modelling
and data.  Changing the adopted mass loss rate affects the HCN
abundance but not, to a first approximation, the HCN envelope size.
Since the HCN lines are mostly radiatively excited, mainly through the
3-$\mu$m stretching mode which lies close to the peak of the spectral
energy distribution, changing the kinetic temperature structure within
reasonable limits also does not significantly affect the model
results.  While the HCN $J=1\rightarrow0$ hyperfine lines can be
inverted in parts of the envelope, the effect of maser emission on the
scales observed here ($\gtrsim 1\arcsec$) should be small.  Previous
modelling of H$^{12}$CN and H$^{13}$CN data for a sample of carbon
stars by \citet{Lindqvist00} indicates that our technique may be less
successful at modelling lines of high optical depth such as HCN, where
changes in the physical parameters in the envelope, such as density
structure and velocity field, will have large effects.  

A higher degree of sophistication in the modelling would require more
detailed observations.  
The near-term prospect of higher resolution HCN $J=1\rightarrow0$
observations is remote, but higher resolution mapping of other
molecules (such as CO and CN) would be useful.  In particular, good
spatial resolution will be important in order to address the
possibility that the medium is highly clumped, as has been suggested
by \citet{Olofsson:96}.


\begin{figure}
\resizebox{\hsize}{!}{\includegraphics{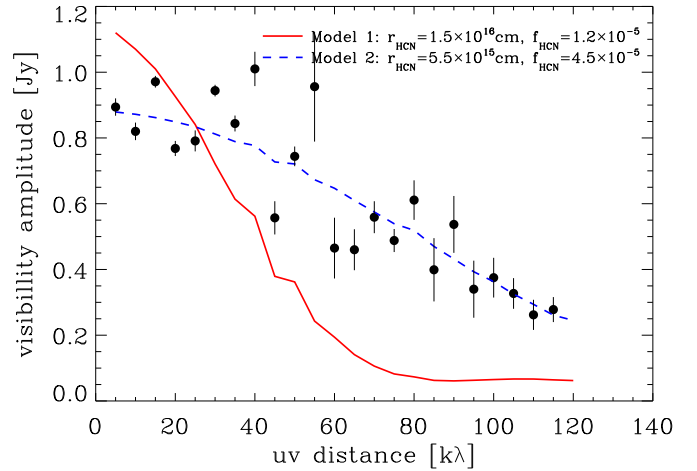}}
\caption{Visibility amplitudes of the observed HCN $J=1\rightarrow0$
line emission obtained at ATCA towards \object{R Scl} as a function of
the projected baseline length.  The observations, integrated from
$-28$ to $-4$~km\,s$^{-1}$ and binned to 5\,k$\lambda$, are plotted as
filled symbols with 1$\sigma$ error bars. Also shown is the result of
applying the same $uv$ sampling to the best-fit circumstellar model
including the single-dish data (Model 1; solid line).  A model with
higher abundance and smaller envelope size is able to reproduce the
interferometer observations much better (Model 2; dashed line).
\label{fig:uvmodel}}
\end{figure}

\section{Discussion: The detached shell}

Since our HCN data do not show direct evidence for a detached shell
around R Scl, it is worth reviewing the evidence that such a shell
exists.  First there are the far-infrared colours of this star, which
are similar to those of stars with clearly detached envelopes
\citep{Olofsson:90}.  In addition, the CO $J=3\rightarrow2$ SEST maps
presented by \citet{Olofsson:96}, with an effective resolution of
about 7\arcsec, show the systematic change in source size with
velocity expected from an expanding spherical shell, as well as
off-centre peaks in the central velocity channels that are indicative
of detachment.  Third, as discussed by \citet{Delgado03a}, attempts to
model the CO data with only a continuous mass loss envelope lead to very
high mass loss rates ($\sim$$10^{-5}$ M$_\odot$~yr$^{-1}$), which
cannot be reconciled with the double-peaked line profile nor produce
the extended brightness distribution of the CO $J=3\rightarrow2$ line.
A high mass loss rate is also inconsistent with the fact that the
star is optically bright.  Finally, the detection of a geometrically
thin dust shell in polarised optical light is also noteworthy
\citep{Delgado03a}, although this shell appears to have a radius
roughly twice as large as that of the proposed CO shell.

The modelling presented in this paper provides indirect support for
the hypothesis that R~Scl has experienced a brief episode of intense
mass loss, leading to the formation of a detached CO-emitting shell.
Our derived HCN abundance of about $1 \times 10^{-5}$ is in excellent
agreement with the estimated photospheric HCN abundance of $\sim
8\times 10^{-6}$ for R~Scl determined by \citet{Olofsson93b}.  It is
also roughly consistent with the abundances of $\sim$$5 \times
10^{-5}$ obtained by \citet{Lindqvist00} for a small sample of carbon
stars.  Since for a given line intensity the derived abundance scales
inversely with the mass loss rate \citep{Olofsson93b}, this argues in
favour of our adopted mass loss rate of $2\times 10^{-7}$
M$_\odot$~yr$^{-1}$, which is two orders of magnitude lower than the
mass loss rate needed to model the CO emission (Lindqvist et al., in
prep.).  Furthermore, as noted already by \citet{Olofsson:96}, the HCN
expansion velocity of 10.5 km~s$^{-1}$ is significantly lower than
that of 16.5 km~s$^{-1}$ measured from the CO lines.  The fact that
the `attached' (present mass loss) envelope expands more slowly than
the detached shell is even more pronounced in sources with old
detached shells, e.g. TT~Cyg \citep{Olofsson:98,Olofsson:00}, where
expansion velocities for the attached envelope as low as 4~km~s$^{-1}$
has been measured.  R~Scl is more reminiscent of U~Cam
\citep{Lindqvist:96,Lindqvist:99}, which has a relatively young
detached shell, and whose attached envelope expands more slowly than
the detached shell but not to the extent seen in sources with older
detached shells.  In the case of U~Cam, however, HCN emission is
detected from both the attached envelope and detached shell, whereas
HCN emission in R~Scl is confined to the former.

For the detached shell to be undetected in HCN, its HCN abundance
there must be dramatically lower than in the attached envelope.
Adopting the best fit CO model from Lindqvist et al.\ (in prep.),
where the shell is located at $r \approx 11\arcsec$ and has a total
mass of $\sim$$10^{-2}$~M$_{\odot}$, an upper limit of
$f_{\mathrm{HCN}} \lesssim 2 \times 10^{-7}$ is obtained.  The SEST
$J=1\rightarrow0$ data provide the best constraint on this value since
a higher HCN abundance in the detached shell would produce more
emission in the line wings ($|V-V_{\star}|\gtrsim 11$ km~s$^{-1}$)
than what is observed.  From the ATCA data alone, we obtain an upper
limit of $f_{\mathrm{HCN}} \lesssim 1 \times 10^{-6}$ in the detached
shell.  The most plausible explanation is the rapid photodissociation
of HCN to form CN as the envelope expands away from the star.  The
difficulty in detecting HCN from sources with larger detached shells
(e.g., S~Sct, U~Ant, and TT~Cyg) is consistent with this hypothesis
\citep{Olofsson:96}.  A more extended distribution of CN as compared
to HCN has been directly observed in circumstellar envelopes around
IRC+10216 \citep{Dayal:95}, U~Cam \citep{Lindqvist:96}, and LP~And
\citep{Lindqvist00}.  \citet{Lindqvist00} show how this phenomenon can
be accounted for by a simple photodissociation model.

\section{Conclusions}

We have presented interferometric observations of HCN $J=1\rightarrow0$
emission around R~Scl that reveal a compact central source with
$FWHM$ of 1\arcsec, although the data do not sample small enough scales
to distinguish a Gaussian from, e.g., a disk or an optically
thin sphere.  We see no clear indication that the source structure
varies with velocity, as would be expected from an expanding spherical
shell, but such variations would be difficult to discern because the
source is only moderately resolved, and because of the hyperfine
structure of the HCN line.  Our interferometric data recover, to
within the calibration uncertainties, all of the single-dish flux
observed with the SEST.

The compact HCN morphology and the narrow line profile of the
higher-transition HCN lines indicate that the HCN emission is
associated with the present mass loss wind.  We estimate a mass loss
rate of $2 \times 10^{-7}$ M$_\odot$~yr$^{-1}$, significantly less
than that needed to account for the detached shell inferred from
single-dish CO data.  This supports the idea that the star has
undergone a recent (within $\sim$$10^3$~yr) episode of intense mass
loss.  The low HCN abundance in the detached shell, which we estimate
to be $f_{\mathrm{HCN}}\lesssim 2\times 10^{-7}$, is consistent with
the rapid photodissociation of HCN into CN as it expands away from the
star.

Modelling the present mass loss envelope using a detailed
radiative transfer code, the ATCA data suggest an HCN abundance and
envelope size of $4.5\times 10^{-5}$ and $5.5\times 10^{15}$~cm,
respectively.  However, such a model cannot be reconciled with
single-dish observations of higher $J$ transitions. The single-dish
constraints are much better met using an HCN abundance of $1.2 \times
10^{-5}$ and an HCN envelope size of $1.4\times 10^{16}$~cm.  The
discrepancy may be due to some limitation in the observational data,
or to deviations from the adopted circumstellar model with regards to
geometry, velocity law, clumpiness, or radiative excitation.  However,
it is not possible within the adopted circumstellar model to reconcile
the interferometer and single-dish data by just changing the mass loss
rate.

These observations demonstrate the imaging potential of
the Australia Telescope at millimetre wavelengths, even at this early
stage in the system development.  Future high-resolution observations
of CN and CO emission from this object, which are likely to reveal a
detached shell morphology, are eagerly anticipated.  Such observations
could be conducted with the fully upgraded ATCA and, at higher
transitions, with the soon-to-be completed Sub-Millimeter Array (SMA).

\begin{acknowledgements}

We thank the many ATNF staff that made these observations possible
through their hard work on the ATCA upgrade.  Special thanks to
M. Kesteven for helping to diagnose and correct the phase errors.  We
also thank S. J. Curran for help with some of the observations.  We
are also grateful to J. H. Bieging for providing us with his HHT
observations.  TW is supported by a Bolton Fellowship at the ATNF.\@
FLS, ML and HO acknowledge financial support from The Swedish Research
Council.  FLS further acknowledges support from the Netherlands
Organization for Scientific Research (NWO) grant 614.041.004.

\end{acknowledgements}

\bibliographystyle{aa}
\bibliography{rscl}

\end{document}